\documentclass[pre,aps,twocolumn,showpacs]{revtex4}
\usepackage{amsmath,amssymb,eucal,graphicx}
\usepackage{epsfig}
\usepackage{textcomp}

\begin{document}
\title{Fluctuations of Current in Non-Stationary Diffusive Lattice Gases}
\author{P. L. Krapivsky}
\affiliation{Physics Department, Boston University, Boston MA 02215, USA}
\author{Baruch Meerson}
\affiliation{Racah Institute of Physics, Hebrew University of
Jerusalem, Jerusalem 91904, Israel}



\begin{abstract}
We employ the macroscopic fluctuation theory to study fluctuations of integrated current in one-dimensional lattice gases with a step-like initial density profile. We analytically determine the variance of the current fluctuations for a class of diffusive processes with a density-independent diffusion coefficient, but otherwise arbitrary. Our calculations rely on a perturbation theory around the noiseless hydrodynamic solution. We consider both quenched and annealed types of averaging (the initial condition is allowed to fluctuate in the latter situation).  The general results for the variance are specialized to a few interesting models including the symmetric exclusion process and the Kipnis-Marchioro-Presutti model. We also probe large deviations of the current for the symmetric exclusion process. This is done by numerically solving the governing equations of the macroscopic fluctuation theory using an efficient iteration algorithm.
\end{abstract}
\maketitle

\section{Introduction}
\label{intro}

Fluctuations around equilibrium states of matter is a classical subject of statistical physics. Close to equilibrium, fluctuations of macroscopic observables are fully described in terms of the free energy \cite{LLStat}. An important recent advance is the elucidation of the behavior of fluctuations, including large deviations, of macroscopic observables in {\em non-equilibrium steady states} (NESS)  of driven lattice gases: simple diffusive transport systems with particle conservation \cite{Spohn,L99,KL99,SZ95,D98,S00,BE07,KRB10}.  The distribution of fluctuations in the NESS, as described by the large deviation functional  \cite{NESS},  can exhibit qualitatively new features, such as non-locality and phase transitions, see Refs. \cite{D07,Jona} for reviews.   So far, large fluctuations in NESS have only been studied in a very few simple lattice gas models. These studies, however, have greatly increased the general understanding of fluctuations around NESS.

Over the last decade a powerful framework, the macroscopic fluctuation theory (MFT) of Bertini, De Sole, Gabrielli, Jona-Lasinio, and Landim \cite{Bertini}, has been developed for the NESS of diffusive lattice gases driven by reservoirs at the boundaries. The MFT is a classical Hamiltonian field theory \cite{Bertini,Tailleur} which describes macroscopic fluctuations in these systems. The MFT formalism is a further development of the low-noise Freidlin-Wentzell theory \cite{FW84} which in turn is a variant of the WKB (after  Wentzel, Kramers and Brillouin) approximation. A celebrated analog of the MFT for continuous stochastic systems is the Martin-Siggia-Rose field-theoretical formalism  \cite{MSR} which has been employed in numerous works.  Related approaches for lattice gases deal, in addition to diffusive transport, with on-site reactions among particles \cite{EK,MS2011,MSfronts}.

The MFT has been successfully applied to NESS in different systems \cite{Bertini,Tailleur,Appert-Rolland,battery,van}, including those driven not  from the boundaries.  Large fluctuations around NESS have also been studied at the microscopic level using exact \cite{DL,van_1,Rakos,KM} and numerical \cite{
kurchan,pablo,
vander} approaches. A perfect agreement between the predictions of the MFT and the long-time asymptotes of the microscopic calculations has been observed whenever the results of the two approaches were available.

With the continuing progress in the studies of NESS, a natural next step is to probe fluctuations of macroscopic observables around \emph{non-stationary} states.  Fortunately, the MFT framework is readily extendable to non-stationary settings, such as evolution of a step-like initial density profile \cite{DG2009b}. There is, however, a major technical hurdle which slows down the progress in using the MFT for the analysis of both stationary and non-stationary problems. Already in the simplest setting of a single species of particles, the MFT involves two coupled nonlinear partial differential equations: the field-theoretical Hamilton equations for the density field (a ``coordinate") and a conjugate  momentum field. With a few exceptions, these equations are not soluble analytically. Still, there are several important factors that make the MFT a viable alternative to other approaches:

\begin{enumerate}
\item{The MFT is stripped off unnecessary details of microscopic interactions, so it directly probes the
large-scale, long-time asymptotic regime that is of most interest.}
\item{The MFT provides the ``optimal path": the density profile history which gives a dominant contribution to the probability of observing, say, a given current.}
\item{The Hamilton equations underlying the MFT can be solved numerically with an iteration algorithm \cite{Stepanov}. Alternatively, a numerical minimization of the mechanical action, intrinsic in the MFT, can be performed \cite{Bunin}. These numerical algorithms are much more computationally efficient than microscopic stochastic simulations.}
\item{As we show here, a perturbative analytic solution is possible which probes, for a whole class of models, small fluctuations.}
\end{enumerate}

The main objective of this work is to demonstrate these advantages.  We will investigate, within the MFT formalism, the noisy evolution of a step-like initial density profile in a class of lattice gas models in one dimension, where the transport is symmetric and diffusion-like.  Two non-trivial examples are the simple symmetric exclusion process (SSEP) which has been extensively studied (see e.g. \cite{Spohn,KL99,L99,SZ95,D98,S00,BE07,KRB10} and references therein) and the Kipnis-Marchioro-Presutti (KMP) model \cite{KMP,BGL,van_2}. (In the SSEP each particle can hop to a neighboring site at rate 1 if that site is unoccupied by another particle. If it is occupied, the move is forbidden. The KMP model is a one-dimensional chain of mechanically uncoupled harmonic oscillators which randomly redistribute energy among neighbors.) The step-like initial condition for the particle density,
\begin{equation}
\label{step}
\rho(x,t=0) =
\begin{cases}
\rho_-,   & x<0,\\
\rho_+,  & x>0,
\end{cases}
\end{equation}
provides a good ``litmus test" for theory of fluctuations in non-stationary systems. As in Refs. \cite{DG2009b,DG2009a}, we will be interested in  the statistics of integrated current -- the total number of particles or the total energy --- passing into the half-line $x>0$ during a given time $T$. The precise mathematical formulation of the problem in the framework of the MFT was  given a few years ago \cite{DG2009b}, but the problem has defied solution except for the completely integrable case of non-interacting random walkers. Our strategy here will be to solve the  problem perturbatively around the noiseless hydrodynamic solution, thus probing typical, small fluctuations of the current. In addition, we will show how large current fluctuations can be efficiently simulated numerically.

We will consider a class of models whose hydrodynamic description is provided by a diffusion equation
\begin{equation}
\label{rho:eq}
\partial_t \rho = \partial_x \!\left[D(\rho) \,\partial_x \rho\right]
\end{equation}
with the diffusion coefficient $D(\rho)$. Having solved this equation with the initial condition \eqref{step}, one can compute
\begin{equation}
\label{current}
\langle J(T)\rangle = \int_0^\infty dx\, [\rho(x,T)-\rho(x,0)],
\end{equation}
the \emph{average} integrated current for the underlying microscopic model. At the level of MFT, the class of microscopic models that we consider here is fully characterized, in addition to the diffusion coefficient $D(\rho)$,  by the function $\sigma(\rho)$ which describes equilibrium fluctuations \cite{Spohn}. Our formalism can handle, in a simple way, systems with a density-independent diffusion coefficient (which we set to $D=1$ without loss of generality) but an arbitrary $\sigma(\rho)$.

The total current $J=J(T)$ into the right half-line is a random quantity. The average total current grows as $\sqrt{T}$ in the long time limit, $T\gg 1$. The variance  of the total current, $\langle J^2\rangle_c=\langle J^2\rangle-\langle J\rangle^2$, also exhibits a diffusive growth with time:
\begin{equation}
\label{var_asympt}
\langle J^2\rangle_c=V(\rho_-,\rho_+, \sigma)\,\sqrt{T}.
\end{equation}
The quantity $V$ depends on the densities $\rho_\pm$ and, through $\sigma=\sigma(\rho)$, on the model. Intriguingly, one has to be careful in defining the averaging procedure \cite{PS02,DG2009b}. In the \emph{quenched} setting the initial condition~\eqref{step} is \emph{deterministic}. In the \emph{annealed} setting one allows equilibrium fluctuations in the initial condition \eqref{step}. More precisely, the initial density profile  in the left (correspondingly, right) part of the system is chosen from the equilibrium
probability distribution corresponding to density $\rho_-$ (correspondingly, $\rho_+$). As a result, \emph{the most probable} initial density profile, see below, is different from a step function.

The main analytical results of this work are explicit expressions for $V$ for diffusive processes with $D=1$ and arbitrary $\sigma(\rho)$. In the quenched setting we obtain
\begin{equation}
\label{variance_quenched}
V_\text{quenched}= \int_0^1 \frac{dt}{4\pi t} \int_{-\infty}^\infty dx\,\sigma[\rho(x,1-t)]\,
e^{-x^2/2t},
\end{equation}
where $\rho(x,t)$ is the solution of the classical diffusion equation with $D=1$ and initial condition \eqref{step}, see Eq.~(\ref{q_0}) below. In the annealed setting we obtain
\begin{equation}
\label{variance_annealed}
V_{\text{annealed}} = V_{\text{quenched}} + \frac{\sqrt{2}-1}{{2\sqrt{2\pi}}}\left[\sigma(\rho_-) + \sigma(\rho_+)\right].
\end{equation}
Since $\sigma(\rho)$ is intrinsically positive (for $\rho>0$), the second term on the right-hand side of Eq.~(\ref{variance_annealed}) is positive. Hence $V_{\text{annealed}} > V_{\text{quenched}}$, as expected on physical grounds.

The rest of this paper is organized as follows. Section \ref{prelim} includes important preliminaries that will be used in the subsequent sections: We briefly discuss the moment generating function of the current and its long-time behavior, formally introduce the functions $D(\rho)$ and $\sigma(\rho)$, and outline the MFT formulation, due to Derrida and Gerschenfeld \cite{DG2009b}, of the problem of statistics of the current for the step-like initial condition (\ref{step}). In Sections~\ref{var} and \ref{var_ann} we develop a perturbation theory around the noiseless hydrodynamic solution and determine the variance of current, alongside with the optimal paths, in the quenched and annealed settings. Particular examples of these results for the symmetric state, $\rho_-=\rho_+$, for the SSEP and KMP models, and for the non-interacting random walkers, are presented in Section~\ref{examples}. Section~\ref{numerics} is devoted to a numerical calculation, within the MFT, of the optimal path conditioned on observing a large deviation of the current. Concluding remarks appear in Sec.~\ref{concl}.
Finally, in one of the Appendices we present an alternative way of calculating the variance in the quenched setting: by employing fluctuating hydrodynamics.

\section{Preliminaries and Governing Equations}
\label{prelim}

\subsection{Moment generating function}
\label{MGF}

A complete description of the current fluctuations is provided by the probability distribution $P(J,T)$. Often it is more convenient to deal with the moment generating function
\begin{equation}
\label{GF}
\left\langle e^{\lambda J} \right\rangle = \sum_{J\geq 0} e^{\lambda J} P(J,T) \equiv
1 + \sum_{n\geq 1} \frac{\lambda^n}{n!}\,\langle J^n\rangle
\end{equation}
that encapsulates $P(J,T)$ and provides the moments of the distribution. Alternatively,
by taking the logarithm of \eqref{GF}, one can rewrite this expression as
\begin{equation}
\label{cumulants}
\ln \left\langle e^{\lambda J} \right\rangle = \sum_{n\geq 1} \frac{\lambda^n}{n!}\,\langle J^n\rangle_c\,,
\end{equation}
which defines the cumulants of the distribution. The first two cumulants are
$\langle J\rangle_c = \langle J\rangle$ and $\langle J^2\rangle_c = \langle J^2\rangle - \langle J\rangle^2$.

In diffusive systems with the step-like initial condition (\ref{step}), the moment generating function exhibits the following long-time behavior:
\begin{equation}
\label{asymp}
\left\langle e^{\lambda J} \right\rangle \sim e^{\sqrt{T}\,\mu(\lambda, \rho_-,\rho_+)}.
\end{equation}
For non-interacting random walkers (RWs), the function $\mu(\lambda,\rho_-,\rho_+)$  was calculated \cite{DG2009b}, both in the quenched and in the annealed settings, from the MFT formalism. For the SSEP it was also calculated \cite{DG2009a}, from the microscopic model,  in the annealed setting:
\begin{equation}
\label{mu_SSEP}
\mu_{\text{annealed}}^{\text{SSEP}} = \frac{1}{\pi}\int_{-\infty}^\infty dk\,\ln\left(1+\Lambda e^{-k^2}\right),
\end{equation}
where
\begin{equation*}
\label{Lambda_SSEP}
\Lambda \!=\! \rho_-(e^\lambda -1) \! +\!  \rho_+(e^{-\lambda} -1) \! + \! \rho_- \rho_+(e^\lambda -1)\,(e^{-\lambda} -1).
\end{equation*}
In the special case of $\rho_-=1$ and $\rho_+=0$, the initial condition in the SSEP cannot fluctuate. As a result, $\mu(\lambda,1,0)$ is the same for both annealed and quenched settings.

\begin{figure}[ht]
\includegraphics[width=2.6 in,clip=]{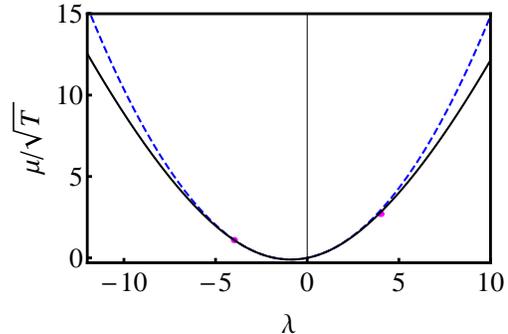}
\caption{(Color online). Plotted versus $\lambda$ are the exact long time result for $\mu_{\text{annealed}}^{\text{SSEP}}/\sqrt{T}$
from Eq.~(\ref{mu_SSEP}) (the solid curve) and the two-cumulant approximation $C_1 \lambda+ (1/2)\, C_2 \lambda^2$ with $C_1$ and $C_2$ from Eqs.~(\ref{C1SSEP}) and (\ref{C2SSEP}) (the dashed curve) for  $\rho_{-}=0.6$ and $\rho_{+}=0.2$. The two circles are numerical results obtained with iteration algorithm described in Section \ref{numerics}.}
\label{muann}
\end{figure}

Expanding the integrand in Eq.~(\ref{mu_SSEP}) in powers of $\lambda$, we can extract the cumulants for the SSEP. They have a universal long-time behavior
\begin{equation}
\label{cumul}
\langle J^p\rangle_c = C_p(\rho_-,\rho_+)\,\sqrt{T},
\end{equation}
with
\begin{eqnarray}
  \sqrt{\pi}\,C_1  &=&  \rho_- - \rho_+,\label{C1SSEP} \\
  \sqrt{\pi}\,C_2  &=& \rho_-+\rho_+ -\rho_-^2-\rho_+^2\nonumber \\
  &+&\left(1-\tfrac{1}{\sqrt{2}}\right)(\rho_- - \rho_+)^2,
\label{C2SSEP}
\end{eqnarray}
etc. Although obtained via an expansion in small $\lambda$, the cumulants $C_1$ and $C_2$ provide a surprisingly good approximation of $\mu$
for $|\lambda|$ comparable to or even greater than 1.  As an example, Fig.~\ref{muann} shows two plots for $\mu_{\text{annealed}}^{\text{SSEP}}/\sqrt{T}$ versus $\lambda$: the exact long-time
result from Eq.~(\ref{mu_SSEP}) and the two-cumulant approximation $\mu=C_1 \lambda+ (1/2)\, C_2 \lambda^2$, for $\rho_{-}=0.6$ and $\rho_{+}=0.2$. As one can see, a discrepancy appears only at $|\lambda|\simeq 5$. Correspondingly, deviations of the probability distribution $P(J,T)$ from gaussianity only occur in far tails of the distribution.

Derrida and Gerschenfeld \cite{DG2009b} also found the function $\mu_{\text{annealed}}(\lambda,\rho_-,\rho_+)$ for the KMP model.   They showed, within the MFT formalism, that it is related to $\mu_{\text{annealed}}$ for the SSEP, so it can be established without a new calculation:
\begin{equation}
\label{mu_KMP}
\mu_{\text{annealed}}^{\text{KMP}} = -\frac{1}{2\pi}\int_{-\infty}^\infty dk\,\ln\left(1+\Lambda e^{-k^2}\right),
\end{equation}
where
\begin{equation}
\label{Lambda_KMP}
\Lambda = 2\lambda(\rho_+ -\rho_-) - 4\lambda^2 \rho_- \rho_+.
\end{equation}

Expanding the integrand in Eq.~(\ref{mu_KMP}) in powers of $\lambda$, yields Eq.~(\ref{cumul}), with the same $C_1$ as for the SSEP, and
\begin{equation}\label{C2KMP}
    \sqrt{\pi} \, C_2^{\text{KMP}} = 4 \rho_{-}\rho_{+} + \sqrt{2} \left(\rho_{-}-\rho_{+}\right)^2\,.
\end{equation}
Notably, the average current,
\begin{equation}
\label{av_SSEP}
\langle J \rangle = \frac{\rho_- -\rho_+}{\sqrt{\pi}}\,\sqrt{T}
\end{equation}
is the same for the annealed and quenched averages, and for any model with $D(\rho)=1$, including non-interacting random walkers, the SSEP and the KMP model. The variance is already model-dependent, and it also depends on the type of averaging. The above expressions for $C_2$ for the SSEP and KMP models refer to the annealed case. To  our knowledge, in the quenched case even the variances are unknown; they will be in the focus of this work.

\subsection{$D$, $\sigma$ and $F$}
\label{functions}

Here is a brief recap of the formal definitions of the quantities $D(\rho)$ and $\sigma(\rho)$, and of their relation to the free energy density in equilibrium,  $F(\rho)$ \cite{Spohn}. The functions $D(\rho)$ and $\sigma(\rho)$ characterize the flux and its variance, respectively, in a simple stationary setting. Consider a one-dimensional system of a finite (but very large) length $L$ which is in contact with reservoirs of particles (or energy) with density $\rho_-$ on the left and $\rho_+$ on the right. When these densities are close to each other, $\rho_\pm\simeq r$, with
\begin{equation}
\label{rr}
|\rho_+-\rho_-|\ll r,
\end{equation}
the average flux per unit time is proportional to $D(r)$ and the density difference:
\begin{equation}
\label{D_def}
\lim_{t\to\infty} \frac{\langle J\rangle}{t} = \frac{D(r)}{L}\,(\rho_- -\rho_+).
\end{equation}
In its turn, $\sigma(r)$ can be extracted from the growth law of the variance of the flux evaluated at the equilibrium state $\rho_\pm=r$:
\begin{equation}
\label{sigma_def}
\lim_{t\to\infty} \frac{\langle J^2\rangle_c}{t} = \frac{\sigma(r)}{L}.
\end{equation}
Therefore, the quantities $D(r)$ and $\sigma(r)$ characterize small deviations from equilibrium.  The equilibrium origin of $D(r)$ and $\sigma(r)$ is additionally emphasized by the equation
\begin{equation}
\label{FDT}
\frac{d^2 F}{dr^2} = \frac{2D(r)}{\sigma(r)},
\end{equation}
relating $D(r)$ and $\sigma(r)$ to the equilibrium free energy density $F(r)$. Equation \eqref{FDT} follows \cite{Spohn,D07} from the fluctuation-dissipation theorem. It also appears naturally in the MFT formalism, see Appendix A.

Table I lists the functions $D(r)$, $\sigma(r)$ and $F(r)$ for three specific models: the RW,  the SSEP and the KMP.

\begin{table}[t]
\label{Table_models}
\begin{tabular}{|c|c|c|c|}
\hline
 ~Model ~ &  ~$D(r)$ ~ & $\sigma(r)$ & ~$F(r)$  \\
\hline
RW      &  1    &      $2r$             &  $r\ln r - r$\\
\hline
SSEP     &  1     &  ~$2r(1-r)$ ~    &   ~$r\ln r  +(1-r)\ln(1-r)$ ~\\
\hline
KMP     &  1     &   ~$4r^2$ ~     &   ~$-(1/2)\,\ln r $ ~\\
\hline
\end{tabular}
\caption{Functions $D(r), \sigma(r),  F(r)$ for non-interacting random walkers and for two interacting particle systems, the SSEP and the KMP.}
\end{table}

\subsection{MFT formalism}
\label{WKB}

The MFT formalism \cite{Bertini,Tailleur,DG2009b}  describes large deviations of macroscopic quantities in diffusive lattice gases.  Mathematically, one must solve two coupled partial differential equations
\begin{equation}
\label{q:eq}
\partial_t q = \partial_x \left[D(q)\, \partial_x q\right]
-  \partial_x \left[\sigma(q)\, \partial_x p\right]
\end{equation}
and
\begin{equation}
\label{p:eq}
\partial_t p = - D(q) \partial_{xx} p
- \frac{1}{2}\sigma^{\prime}(q)\!\left(\partial_x p\right)^2,
\end{equation}
for the density field $q(x,t)$  and the conjugate momentum field $p(x,t)$. Here and in the following the prime denotes derivative with respect to the argument. Solutions with $p(x,t)=0$ are called relaxation solutions. For the relaxation solutions Eq.~\eqref{p:eq} is satisfied, and Eq.~\eqref{q:eq} reduces to the hydrodynamic equation \eqref{rho:eq}, so $q(x,t)= \rho(x,t)$. Solutions with $p(x,t)\neq 0$ are called activation solutions, here $q(x,t)\neq \rho(x,t)$. Equations~\eqref{q:eq} and \eqref{p:eq} are Hamiltonian,
\begin{equation}
\partial_t q = \delta H/\delta p\,, \quad
\partial_t p = -\delta H/\delta q\,,
\end{equation}
with the Hamiltonian
\begin{equation}
\label{Hamiltonian}
H[q(x,t),p(x,t)]=\int_{-\infty}^\infty dx\,\mathcal{H},
\end{equation}
where
\begin{equation}
\label{Ham}
\mathcal{H}(q,p) = -D(q)\, \partial_x q\, \partial_x p
+\frac{1}{2}\sigma(q)\!\left(\partial_x p\right)^2.
\end{equation}

For a given model, specified by $D$ and $\sigma$, Eqs.~\eqref{q:eq} and \eqref{p:eq} can describe large deviations of different quantities in different settings. The problem of statistics of current  during time $T$, starting from a step-like density profile, is specified by certain boundary conditions in $x$ and $t$. To begin with, by virtue of mass conservation, a given integrated current implies an integral constraint
\begin{equation}
\label{current0}
J  =  \int_0^\infty dx\, [q(x,T)-q(x,0)].
\end{equation}
The boundary conditions in $t$ are different for the quenched and annealed settings. (The term `boundary' emphasizes here that these are conditions at the boundaries of the time interval $[0,T]$.)
In the quenched setting, the initial condition for the density coincides with Eq.~\eqref{step}:
\begin{equation}
\label{q_step}
q(x,t=0) = \rho_-\theta(-x) + \rho_+ \theta(x),
\end{equation}
where $\theta(x)$ is the Heaviside step function. The conjugate momentum is constrained by the condition at $t=T$ \cite{DG2009b}
\begin{equation}
\label{p_step}
p(x,t=T) = \lambda \theta(x),
\end{equation}
where the Lagrangian multiplier $\lambda=\lambda(J)$ is fixed by Eq.~(\ref{current0}). Once $q(x,t)$ and $p(x,t)$ are found for $0\leq t\leq T$, one can calculate \cite{DG2009b}
the function $\mu$ that enters Eq.~(\ref{asymp}) for the long-time asymptote of the moment-generating function:
\begin{eqnarray}
\mu_{\text{quenched}} &=& \lambda \int_0^\infty dx\,[q(x,T)-q(x,0)]\nonumber \\
&-& \frac{1}{2}\int_0^T dt \int_{-\infty}^\infty dx\,\sigma(q) (\partial_x p)^2.
\label{muquenched}
\end{eqnarray}
The first term in $\mu_{\text{quenched}}$ comes from the constraint (\ref{current0}), whereas the second one is equal to $-S(T)$, where $S(T)$ is the mechanical action of the Hamiltonian system~\eqref{q:eq} and \eqref{p:eq}. Indeed,
\begin{equation}
\label{action:def}
S(T) = \int_0^T dt \int_{-\infty}^\infty dx \left(p\,\partial_t q - \mathcal{H}\right).
\end{equation}
Using Eqs.~\eqref{q:eq} and \eqref{Ham} and performing integration by parts in the spatial integral in \eqref{action:def}, one can rewrite the action as
\begin{equation}
\label{action}
S(T) = \frac{1}{2}\int_0^T dt \int_{-\infty}^\infty dx \,\sigma(q) (\partial_x p)^2.
\end{equation}

In the annealed setting, the boundary condition at the final time $t=T$ is again given by Eq.~(\ref{p_step}). The initial condition  is now different from Eq.~(\ref{q_step}), it involves both $q$ and $p$ \cite{DG2009b}:
\begin{equation}
\label{p_init}
p(x,0) = \lambda \theta(x) + 2 \int_{\rho(x,0)}^{q(x,0)} dr\,\frac{D(r)}{\sigma(r)}\,.
\end{equation}
Once $q(x,t)$ and $p(x,t)$ are found, the function $\mu$ can be calculated \cite{DG2009b} from the equation
\begin{eqnarray}
\mu_{\text{annealed}} = & - & 2 \int_{-\infty}^\infty dx \int_{\rho(x,0)}^{q(x,0)}dr\,\frac{D(r)}{\sigma(r)}\,
[q(x,0)-r] \nonumber \\
&+& \lambda \int_0^\infty dx\,[q(x,1)-q(x,0)]\nonumber \\
&-& \frac{1}{2}\int_0^T dt \int_{-\infty}^\infty dx\,\sigma(q) (\partial_x p)^2.
\label{muannealed}
\end{eqnarray}
The second and third terms here are the same as in the quenched setting, except that $q(x,t)$ and $p(x,t)$ are different. The first term is specific to the annealed setting: it describes the cost of creating the \emph{optimal initial condition} for the given value of the current.

\section{Variance in the quenched case}
\label{var}

From now on, we will only consider a class of models where $D=1$ (such as in all three examples in Table 1). Here the governing equations \eqref{q:eq} and \eqref{p:eq} become
\begin{subequations}
\begin{align}
\label{q:SSEP}
\partial_t q &= \partial_{xx} q - \partial_x\left[\sigma(q) \partial_x p\right],\\
\label{p:SSEP}
\partial_t p &= - \partial_{xx} p - \tfrac{1}{2}\sigma^{\prime}(q) (\partial_x p)^2.
\end{align}
\end{subequations}
The exact solution of Eqs.~\eqref{q:SSEP} and \eqref{p:SSEP}, subject to the boundary conditions \eqref{q_step} and \eqref{p_step}, is unknown except for the non-interacting random walkers, when $\sigma(q)$ is proportional to $q$. Fortunately, the variance of current can be found for arbitrary
$\sigma(q)$ by using a perturbation expansion over a hydrodynamic solution $(q_0,p_0)=(\rho,0)$, where
\begin{equation}
\label{q_0}
\rho(x,t) = \frac{\rho_- + \rho_+}{2}+\frac{\rho_+ - \rho_-}{2}\,\text{erf}\!\left(\frac{x}{\sqrt{4t}}\right)
\end{equation}
solves Eq.~\eqref{rho:eq}, with $D=1$,  subject to the initial condition \eqref{q_step}. The Lagrangian multiplier $\lambda$ plays the role of a small parameter in this expansion. As one can justify \textit{a posteriori}, a small $\lambda$ implies a small deviation of the current from its average value $\langle J \rangle$.
We expand
\begin{subequations}
\begin{align}
\label{q_exp}
q &= q_0 + \lambda q_1 + \lambda^2 q_2+\ldots \\
\label{p_exp}
p &= \qquad \lambda p_1 + \lambda^2 p_2+\ldots
\end{align}
\end{subequations}
and plug these expansions into Eqs.~\eqref{q:SSEP} and \eqref{p:SSEP}. In the zeroth order we recover $(q_0,p_0)=(\rho,0)$. The first-order equations are
\begin{subequations}
\begin{align}
\label{q_1}
(\partial_t -\partial_{xx}) q_1 &=  -\partial_x[\sigma(\rho)\partial_x p_1],\\
\label{p_1}
\partial_t p_1 &= - \partial_{xx} p_1.
\end{align}
\end{subequations}
The boundary conditions for $q_1$ and $p_1$ follow from Eqs.~\eqref{q_step} and \eqref{p_step}:
\begin{equation}
\label{BC_1}
q_1(x,t=0) = 0, \quad p_1(x,t=T)=\theta(x).
\end{equation}
Solving the anti-diffusion equation \eqref{p_1} with the boundary condition \eqref{BC_1} for $p_1$, we obtain
\begin{equation}
\label{p_1:sol}
p_1(x,t) = \frac{1}{2}+\frac{1}{2}\,\text{erf}\!\left[\frac{x}{\sqrt{4(T-t)}}\right].
\end{equation}
Now we need to solve Eq.~(\ref{q_1}): a diffusion equation with a known source term. The form of equation suggests to seek $q_1$ as a gradient:
\begin{equation}
\label{q_psi}
q_1 = -\partial_x \psi.
\end{equation}
The potential $\psi$ satisfies the equation
\begin{equation}
\label{psi_eq}
(\partial_t - \partial_{xx})\psi  = F,
\end{equation}
where $F=\sigma(\rho) \partial_x p_1$, and
\begin{equation}
\label{p1_diff_T}
\partial_x p_1  = \frac{1}{\sqrt{4\pi(T-t)}}\, \exp\!\left[-\frac{x^2}{4(T-t)}\right].
\end{equation}
The solution of Eq.~(\ref{psi_eq}) is
\begin{equation*}
\label{psi_sol}
\psi(x,t) =  \int_0^t d\tau \int_{-\infty}^\infty dy \,\frac{F(y,\tau)}{\sqrt{4\pi(t-\tau)}}\,
\exp\!\left[-\frac{(x-y)^2}{4(t-\tau)}\right].
\end{equation*}
In particular,
\begin{eqnarray}
\label{psi_integral}
\psi(0,T) & = &  \int_0^T dt \int_{-\infty}^\infty dx\,\frac{\sigma(\rho) \partial_x p_1}{\sqrt{4\pi(T-t)}}\,
\exp\!\left[-\frac{x^2}{4(T-t)}\right] \nonumber\\
& = & \int_0^T dt  \int_{-\infty}^\infty dx \,\sigma(\rho)(\partial_x p_1)^2.
\end{eqnarray}
The function $\mu$ from Eq.~(\ref{muquenched}) becomes
\begin{eqnarray}
\mu_{\text{quenched}} &=& \lambda \langle J\rangle + \lambda^2 \psi(0,T) -\frac{\lambda^2}{2} \psi(0,T) + \dots \nonumber \\
&=& \lambda \langle J\rangle + \frac{\lambda^2}{2} \psi(0,T) + \mathcal{O}(\lambda^3).
\label{muquenched1}
\end{eqnarray}
Using $\mu=\lambda \langle J\rangle +\frac{1}{2} \lambda^2 \langle J^2\rangle_c+\ldots$ [see Eq.~(\ref{GF})], we extract the variance:
\begin{equation}
\label{variance}
\langle J^2\rangle_c = \psi(0,T)=\int_0^T dt \int_{-\infty}^\infty dx \,\sigma(\rho)(\partial_x p_1)^2.
\end{equation}
This result holds, for the quenched setting, for all diffusion processes with $D(\rho)=1$ and arbitrary $\sigma(\rho)$.

The $T$-dependence of $\langle J^2\rangle_c$ can be easily extracted via the transformation $t\to t/T$ and $x\to x/\sqrt{T}$ which reduces Eq.~\eqref{variance} to
\begin{equation}
\label{variance_new}
\langle J^2\rangle_c = V \sqrt{T}, \quad V =  \int_0^1 dt \int_{-\infty}^\infty dx\,\sigma(\rho)(\partial_x p_1)^2.
\end{equation}
Here $\rho(x,t)$ is still given by \eqref{q_0}, whereas $\partial_x p_1$ is obtained from \eqref{p1_diff_T} by setting $T=1$. Plugging these expressions into \eqref{variance_new} yields the announced result \eqref{variance_quenched}.

The variance $\langle J^2\rangle_c$ corresponds to a Gaussian asymptotic of the current distribution:
\begin{equation}\label{gauss2}
    P(J,T)\simeq \frac{1}{\sqrt{2 \pi \langle J^2\rangle_c}}\exp\left[-\frac{\left(J-\langle J \rangle\right)^2}{2\langle J^2\rangle_c}\right].
\end{equation}

One can see from Eq.~(\ref{variance_new}) that the variance $\langle J^2\rangle_c$ depends on the model only through $\sigma(\rho)$. In simple models $\sigma(\rho)$ is a low-degree polynomial, see Table I. Consider now a more general case when $\sigma(\rho)$ admits a representation
\begin{equation}
\label{sigma_series}
\sigma(\rho) = \sum_{n\geq 0} A_n \rho^n
\end{equation}
(The zeroth term in the series \eqref{sigma_series} vanishes, $A_0=0$, since $\sigma(0)=0$.) Combining \eqref{q_0} and \eqref{sigma_series} we get
\begin{equation}
\label{V}
V_\text{quenched} = \frac{1}{4\pi}\sum_{n\geq p\geq 0} \frac{A_n}{2^n}\binom{n}{p} d^p s^{n-p} E_p
\end{equation}
where $d = \rho_+ - \rho_-$, $s = \rho_+ + \rho_-$, and
\begin{equation*}
E_p = \int_0^1 \frac{dt}{t} \int_{-\infty}^\infty dx\, \exp\!\left(-\frac{x^2}{2t}\right)
\left[\text{erf}\!\left(\frac{x}{\sqrt{4(1-t)}}\right)\right]^p.
\end{equation*}
The spatial integrals in $E_p$ vanish when $p$ is odd, while for even $p$ one finds $E_0 = \sqrt{8\pi}\,, E_2 = \left(3-2\sqrt{2}\right)\sqrt{8\pi}$, \textit{etc}. The knowledge of $E_p$ with $p\leq 3$ suffices to determine the variance for the 3-parameter class of models
\begin{equation*}
\sigma = A_1\rho + A_2\rho^2 + A_3\rho^3.
\end{equation*}
Here one obtains
\begin{eqnarray}
\label{AAA}
V_\text{quenched} &=& \frac{1}{8\sqrt{2\pi}}\left(4A_1\,s+2A_2\,s^2+A_3\,s^3\right) \nonumber\\
& + & \frac{3-2\sqrt{2}}{8\sqrt{2\pi}}\left(2A_2\,d^2+3A_3\,d^2 s\right).
\end{eqnarray}

We have also calculated the variance in the quenched setting from fluctuating hydrodynamics, see
Appendix B.

\section{Variance in the Annealed Case}
\label{var_ann}

In the annealed case, the calculations are very similar, albeit somewhat more cumbersome. Employing the same perturbation expansion \eqref{q_exp} and \eqref{p_exp}, we recast the initial condition \eqref{p_init} into
\begin{equation*}
p(x,0) = \lambda \left[\theta(x) + \frac{2q_1(x,0)}{\sigma[\rho(x,0)]}\right]+\mathcal{O}(\lambda^2),
\end{equation*}
which yields
\begin{equation}
\label{p1_1}
p_1(x,0) = \theta(x) + \frac{2q_1(x,0)}{\sigma[\rho(x,0)]}.
\end{equation}
On the other hand, Eq.~\eqref{p_1:sol} is still valid, and therefore
\begin{equation}
\label{p1_2}
p_1(x,0) = \frac{1}{2}+\frac{1}{2}\,\mathcal{E}(x), \quad
\mathcal{E}(x)\equiv \text{erf}\!\left(\frac{x}{2}\right).
\end{equation}
(The $T$-dependence here is the same as in the quenched case, so we set $T=1$.)
Comparing Eqs.~\eqref{p1_1} and \eqref{p1_2}, we can deduce the initial condition on $q_1$:
\begin{equation}
\label{q1_init}
q_1(x,0) = \frac{1}{4}\times
\begin{cases}
\sigma(\rho_-) \left[\mathcal{E}(x)+1\right],  & x < 0,\\
\sigma(\rho_+) \left[\mathcal{E}(x)-1\right],  & x>0.
\end{cases}
\end{equation}
As in the quenched case, we employ a gradient representation \eqref{q_psi} and find that the potential $\psi$ satisfies the same inhomogeneous diffusion equation \eqref{psi_eq}. In the quenched case we had $q_1(x,0)=0$ that led to $\psi(x,0)=0$. In the annealed case the initial condition
\eqref{q1_init} leads to a non-trivial initial condition for $\psi$:
\begin{eqnarray}
\label{psi_init}
\!\psi(x,0) &=& \!\theta(-x)\,\sigma(\rho_-)
\left[\frac{1-e^{-x^2/4}}{2\sqrt{\pi}}- x\frac{\mathcal{E}(x)+1}{4}\right] \nonumber\\
\!&+& \!\theta(x)\,\sigma(\rho_+)\!
\left[\frac{1-e^{-x^2/4}}{2\sqrt{\pi}}- x\frac{\mathcal{E}(x)-1}{4}\right],
\end{eqnarray}
where we have demanded that $\psi(x,0)$ be continuous at $x=0$ and chosen the arbitrary constant so that $\psi(0,0)=0$. We now plug the $\lambda$-expansions into Eq.~(\ref{muannealed}) for $\mu_{\text{annealed}}$ and obtain
\begin{eqnarray}
\label{mu_ann_exact}
\!\!\mu_{\text{annealed}} &=& -\lambda^2 \int_{-\infty}^\infty dx\, \frac{[q_1(x,0)]^2}{\sigma[\rho(x,0)]} \nonumber  \\
\!\!&+& \lambda \langle J\rangle +\lambda^2 \psi(0,1) \nonumber  \\
\!\!&-& \frac{\lambda^2}{2}\int_0^1 dt \int_{-\infty}^\infty dx\,\sigma(\rho) (\partial_x p_1)^2+\dots
\end{eqnarray}
with the same average current $\langle J\rangle$ as in the quenched setting. The variance is again extracted by using the expansion $\mu=\lambda \langle J\rangle +\frac{1}{2} \lambda^2 \langle J^2\rangle_c+\ldots$. The result is
\begin{eqnarray*}
V_{\text{annealed}} = & - &\frac{1}{8} \int_{-\infty}^0 dx\, \sigma(\rho_-) \left[\mathcal{E}(x)+1\right]^2\\
&-&\frac{1}{8} \int_0^\infty dx\, \sigma(\rho_+) \left[\mathcal{E}(x)-1\right]^2 \\
&+& 2 \psi(0,1)- \int_0^1 dt \int_{-\infty}^\infty dx\,\sigma(\rho) (\partial_x p_1)^2.
\end{eqnarray*}
After some algebra we find
\begin{eqnarray*}
\label{psi_01}
\psi(0,1) &=&  \int_0^1 dt  \int_{-\infty}^\infty dx\,\sigma(\rho)(\partial_x p_1)^2\nonumber\\
&+& \int_{-\infty}^\infty dx\,\frac{\psi(x,0)}{\sqrt{4\pi}}\,e^{-x^2/4},
\end{eqnarray*}
where $\psi(x,0)$ is given by \eqref{psi_init}.  Combining everything and evaluating integrals, we arrive at the announced result \eqref{variance_annealed} for the variance.

\section{Examples}
\label{examples}

We now specialize the results to the three well-known models presented in Table 1. Prior to that, however, we consider, for an arbitrary $\sigma(\rho)$, the symmetric case $\rho_-=\rho_+=\rho$.

\subsection{Symmetric case}

For $\rho_-=\rho_+=\rho$ the expression $\sigma(\rho)$ can be taken out of the integral in Eqs.~\eqref{variance_quenched} and \eqref{variance_annealed}, and we arrive at
\begin{subequations}
\begin{align}
\label{var_gen_q}
V_\text{quenched}  &= \frac{\sigma(\rho)}{\sqrt{2\pi}}\,,\\
\label{var_gen_a}
V_\text{annealed}  &= \frac{\sigma(\rho)}{\sqrt{\pi}}\,.
\end{align}
\end{subequations}
Hence in the symmetric case $V_\text{annealed} = \sqrt{2}\, V_\text{quenched}$ independently of $\rho$ and for arbitrary $\sigma(\rho)$. In the particular case of SSEP, Eq.~(\ref{var_gen_a}) has been known for a long time, see Ref. \cite{varadhan} and references therein, whereas  Eq.~(\ref{var_gen_q}) (again, for the SSEP) has been obtained only recently \cite{varadhan}. Derrida and Gerschenfeld \cite{DG2009b} noticed that, in models obeying the particle-hole symmetry, there is a symmetry relation between the optimal profiles in the quenched and annealed cases. For the SSEP, this relation leads to
\begin{equation}
\mu_{\text{annealed}}(\lambda,1/2)=\sqrt{2}\, \mu_{\text{quenched}}(\lambda,1/2),
\end{equation}
for $\rho_-=\rho_+=1/2$ \cite{DG2009b}. We emphasize that relation $V_\text{annealed} = \sqrt{2}\, V_\text{quenched}$, which follows from Eqs.~(\ref{var_gen_q}) and  (\ref{var_gen_a}), is valid for any $\rho$ and it does not require the particle-hole symmetry, although we have derived it for only the variances.

\subsection{RWs}

For random walkers the variance is linear in the densities $\rho_-$ and $\rho_+$:
\begin{equation}
\label{RW-var}
V_\text{quenched} = \frac{\rho_+ + \rho_-}{\sqrt{2\pi}}\,, \quad
\frac{V_\text{annealed}}{V_\text{quenched}} = \sqrt{2},
\end{equation}
in agreement with Ref.~\cite{DG2009b}.

\subsection{SSEP}

Specializing Eq.~\eqref{AAA} to $A_1=2, A_2=-2$, and $A_3=0$, we find the variance for the SSEP in the quenched setting:
\begin{eqnarray}
\sqrt{2\pi}\,V_\text{quenched} &=& \rho_+ + \rho_- - \frac{(\rho_+ +\rho_-)^2}{2} \nonumber \\
& - & \frac{3-2\sqrt{2}}{2}(\rho_+ -\rho_-)^2.
\label{asymfullquen}
\end{eqnarray}
To our knowledge, this result is new. In the annealed setting we get
\begin{eqnarray}
\sqrt{\pi}\, V_\text{annealed} &=& \rho_-+\rho_+ -\rho_-^2-\rho_+^2 \nonumber \\
&+& \left(1-\tfrac{1}{\sqrt{2}}\right)(\rho_- - \rho_+)^2,
\label{asymfull}
\end{eqnarray}
in agreement with Eq.~(\ref{C2SSEP}).  In the extreme anti-symmetric case $(\rho_-,\rho_+)=(\rho,0)$
the variance reads
\begin{equation}
\label{asym}
\begin{split}
V_\text{quenched} &= \frac{\rho}{\sqrt{2\pi}}-\frac{2-\sqrt{2}}{\sqrt{2\pi}}\,\rho^2,\\
V_\text{annealed} &= \frac{\rho}{\sqrt{\pi}}-\frac{\rho^2}{\sqrt{2\pi}}.
\end{split}
\end{equation}
One can see that $V_\text{quenched}<V_\text{annealed}$ for all $0<\rho<1$; the equality occurs only when $\rho=0$ or $\rho=1$.

\subsection{KMP}

For the KMP model we have $A_2=4$ and $A_1 = A_3 = 0$. As a result,
\begin{eqnarray*}
 \!\! V_\text{quenched} &=& \frac{(\rho_+ +\rho_-)^2 + (3-2\sqrt{2})(\rho_+ -\rho_-)^2}{\sqrt{2\pi}}\,,
  \label{KMPq}\\
 \!\! V_\text{annealed} &=& \frac{4 \rho_- \rho_+}{\sqrt{\pi}} + \frac{2(\rho_- -\rho_+)^2}{\sqrt{2\pi}}\,.
  \label{KMPa}
\end{eqnarray*}
The expression for $V_{\text{annealed}}$ coincides with $C_2$ from Eq.~(\ref{C2KMP}), the expression for $V_{\text{quenched}}$
is new.

\section{Large deviations: a numerical solution}
\label{numerics}

What happens when $\lambda$ is not small or, in other words, the current $J(T)$ is not close to the average current $\langle J\rangle$? As we cannot solve the MFT equations (\ref{q:eq}) and (\ref{p:eq}) analytically, we resort to a numerical solution. Bunin \textit{et al.} \cite{Bunin} have recently developed
a numerical algorithm based on minimization of the mechanical action. In their algorithm, the boundary conditions in time involve the knowledge of the density profiles at some initial and final times. In the context of integrated current fluctuations, one needs an algorithm that would deal with a boundary condition on the \emph{momentum} at $t=T$.  Fortunately,  classical field-theoretic Hamilton equations have previously appeared in many different contexts. It is hardly surprising, therefore, that an efficient and simple numerical algorithm, with the required type of boundary condition at $t=T$, already exists. It was originally suggested by Chernykh and Stepanov \cite{Stepanov} for evaluating the probability distribution of large negative velocity gradients in the Burgers turbulence.  Later on it was employed by Elgart and Kamenev \cite{EK} and Meerson and Sasorov \cite{MS2011} for evaluating the mean time to extinction in finite-size lattice gas systems involving random walk  and on-site reactions.

The algorithm iterates the diffusion-type Eq.~(\ref{q:eq}) forward in time and the anti-diffusion-type Eq.~(\ref{p:eq}) backward in time.  Consider first the quenched case. In a simple version of the algorithm, each iteration of $q(x,t)$ starts at $t=0$ from the initial condition~(\ref{q_step}) and solves Eq.~(\ref{q:eq}) forward in time until time $t=1$ is reached. In this calculation the previous iteration for $p(x,t)$ is used.  Then Eq.~(\ref{p:eq}) for $p$ is solved backward in time starting, at $t=1$, from $p(x,1)=\lambda \theta(x)$ [see Eq.~(\ref{p_step})], and continuing until  $t=0$. Here the previous iteration for $q(x,t)$ is used. The very first iteration for $p$ is simply the desired final state $p(x,1)=\lambda \theta(x)$.

Unfortunately, the simple version of the algorithm suffers from a numerical instability: after an initial transient, the numerical solution alternates between two different sets of $q$ and $p$, instead of converging to a unique $(q,p)$ solution \cite{previous}. Similarly to Ref.~\cite{Stepanov}, we suppressed this instability by replacing $p$, in the iterations for $q$, by a linear combination of the values of $p$ obtained in \emph{two} previous iterations.  Similarly, we replaced $q$, in the iterations for $p$, by the same linear combination of the values of $q$ obtained in two previous iterations. The relative weights of these two values of $p$ and $q$ (the coefficients of the linear combination) must sum up to unity. In the examples shown below we chose the previous iteration with weight $0.75$ and the iteration before previous with weight $0.25$. The first two iterations of $q$ and $p$ are performed with the simple version of the algorithm.

For the annealed setting we have to use the initial condition~(\ref{p_init}) which involves both $q$ and $p$ which are a priori unknown. Therefore, in the very first iteration we solve Eq.~(\ref{q:eq}) for $q$ forward in time, starting from a ``wrong" (quenched) initial condition,  Eq.~(\ref{q_step}). Then, after solving Eq.~(\ref{p:eq}) for $p$  backward in time until $t=0$, we determine $q(x,0)$ from Eq.~(\ref{p_init}), feed it into
the forward-in-time solution for $q$, and continue iterations. Here too we used, starting from the third iteration, the values from \emph{two} previous iterations to suppress the numerical instability.

\begin{figure}[ht]
\includegraphics[width=2.4 in,clip=]{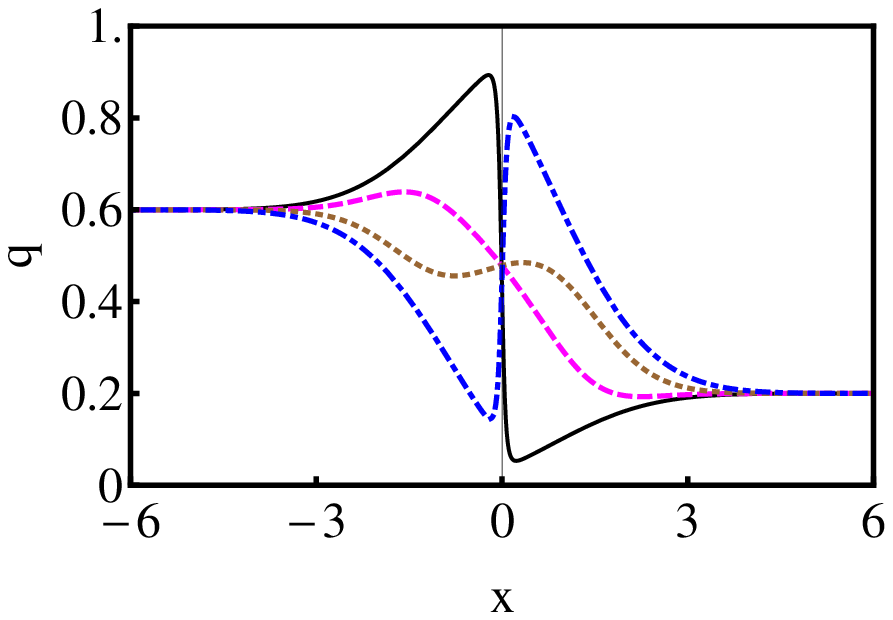}
\includegraphics[width=2.2 in,clip=]{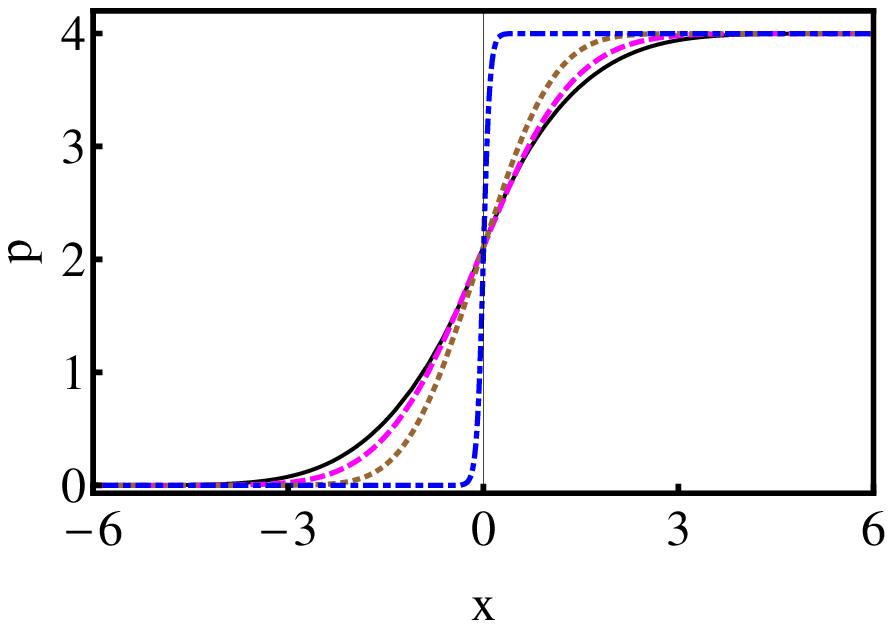}
\caption{(Color online) A numerically computed optimal path for the SSEP with a step-like initial density
profile ($\rho_-=0.6$, $\rho_+=0.2$) in the annealed setting. Shown are the particle density $q(x,t)$ (upper panel) and the conjugate momentum density $p(x,t)$ (lower panel) at rescaled time moments $t=0$ (solid line), $1/3$ (dashed line), $2/3$ (dotted line) and $1$ (dash-dotted line). The Lagrangian multiplier $\lambda=4$ corresponds to the rescaled current $J=\mu^{\prime}(\lambda)|_{\lambda=4}= 1.118\dots$, where $\mu(\lambda)$ is taken
from Eq.~(\ref{mu_SSEP}). For comparison, the rescaled average current is $\langle J\rangle=(\rho_--\rho_+)/\sqrt{\pi}=0.2256\dots$.}
\label{annealedqp4}
\end{figure}

\begin{figure}[ht]
\includegraphics[width=2.4 in,clip=]{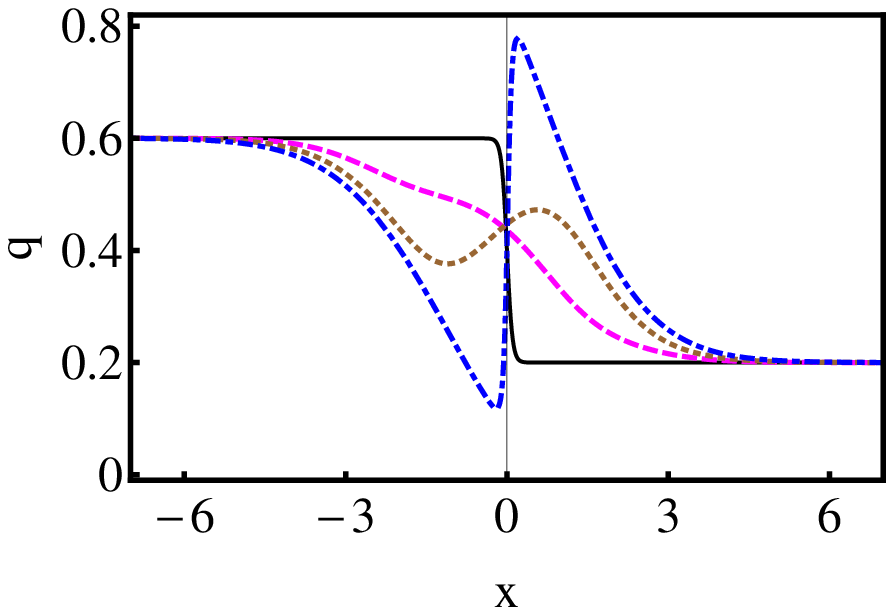}
\includegraphics[width=2.3 in,clip=]{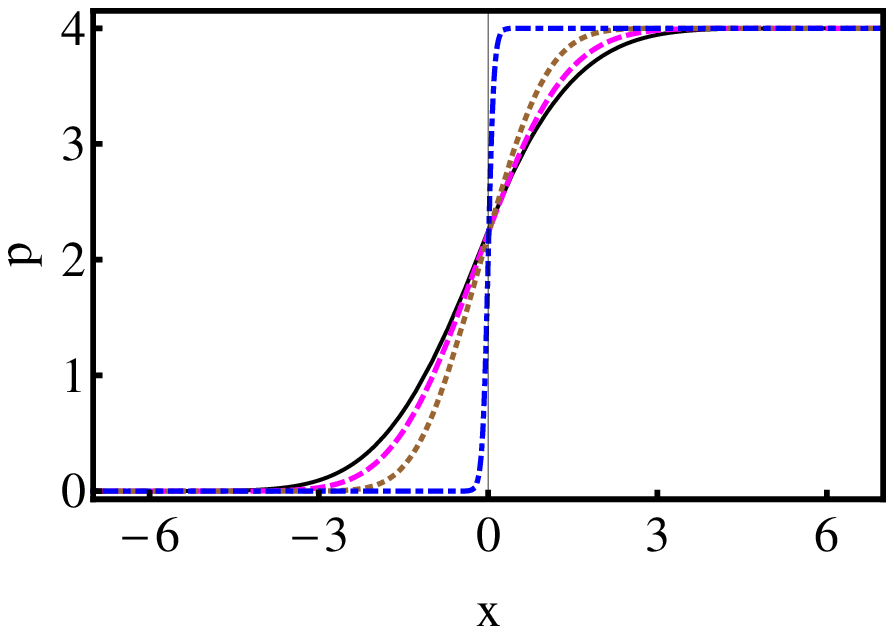}
\caption{(Color online) Same as in Fig. \ref{annealedqp4}, but in the quenched setting. Here $\lambda=4$ corresponds to the
rescaled current $J\simeq 0.9$, as found numerically.}
\label{quenchedqp4}
\end{figure}

We implemented this algorithm in {\it Mathematica}. We worked with a finite-size system, $|x|<L/2$ and  imposed the boundary conditions  $q(-L/2,t)=\rho_-$, $q(L/2,t)=\rho_+$, $p(-L/2,t)=0$, and $p(L/2,t)=\lambda$. The step-functions entering the boundary conditions at $t=0$ and $t=1$ were smoothed a bit. The iterations converge very rapidly. Having computed $q(x,t)$ and $p(x,t)$, one can evaluate the large-deviation function $\mu$ by numerically evaluating the integrals in Eqs.~(\ref{muquenched}) and (\ref{muannealed}) for the quenched and annealed settings, respectively.

Figure \ref{annealedqp4} shows an example of numerically found optimal path $q(x,t)$, and the corresponding $p(x,t)$, for the SSEP in the annealed setting. In this case the function $\mu$ is known, see Eq.~(\ref{mu_SSEP}).
$\lambda=4$ corresponds to a positive current about five times greater than the average current. The optimal initial profile is such as to facilitate \emph{hydrodynamic} transport that does not cost
action. The  optimal fluctuation grows towards $t=1$, when the hydrodynamic flow weakens. In this example our numerically computed values of the function $\mu$ and rescaled integrated current [using Eqs.~(\ref{muannealed}) and (\ref{current}), respectively] are within 3 and 4 per cent, respectively, of their theoretical values. The numerically found values of $\mu$ for $\lambda=4$ and $-4$ are shown in Fig.~\ref{muann} alongside with the exact and approximate analytical results for $\mu(\lambda)$.

\begin{figure}[ht]
\includegraphics[width=2.4 in,clip=]{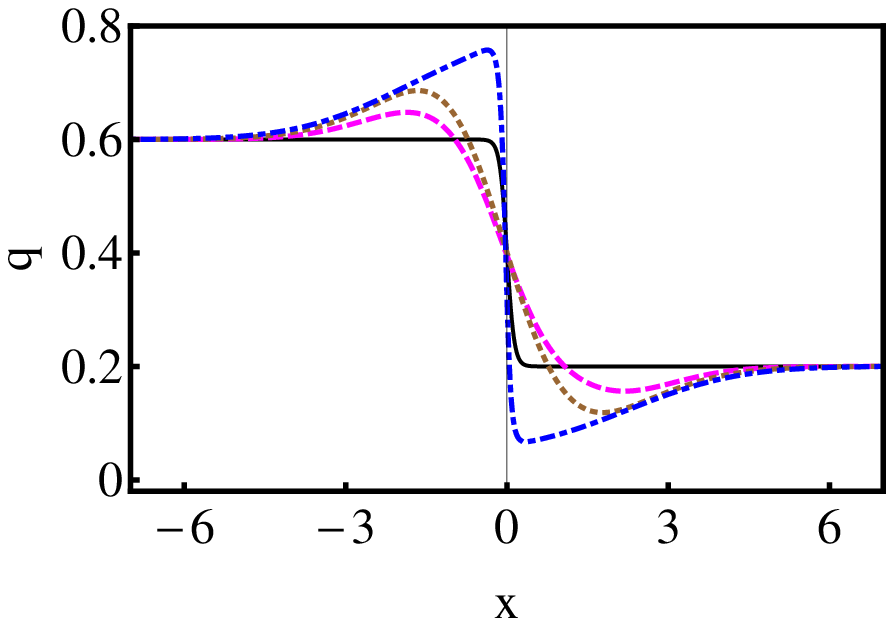}
\includegraphics[width=2.3 in,clip=]{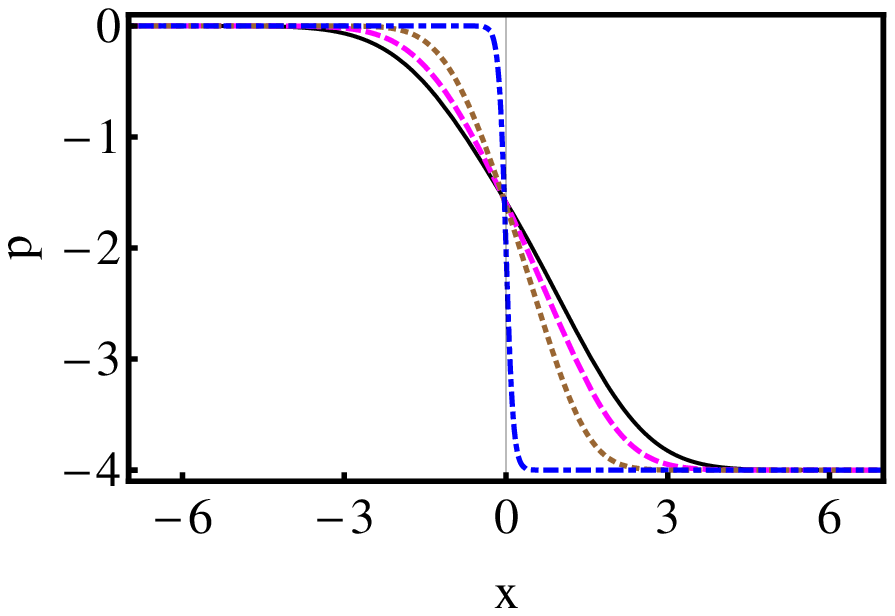}
\caption{(Color online) Same as in Fig. \ref{quenchedqp4}, but with $\lambda=-4$ which corresponds
to a negative rescaled current $J\simeq -0.3$, as found numerically.}
\label{qpminus4}
\end{figure}

\begin{figure}[ht]
\includegraphics[width=2.4 in,clip=]{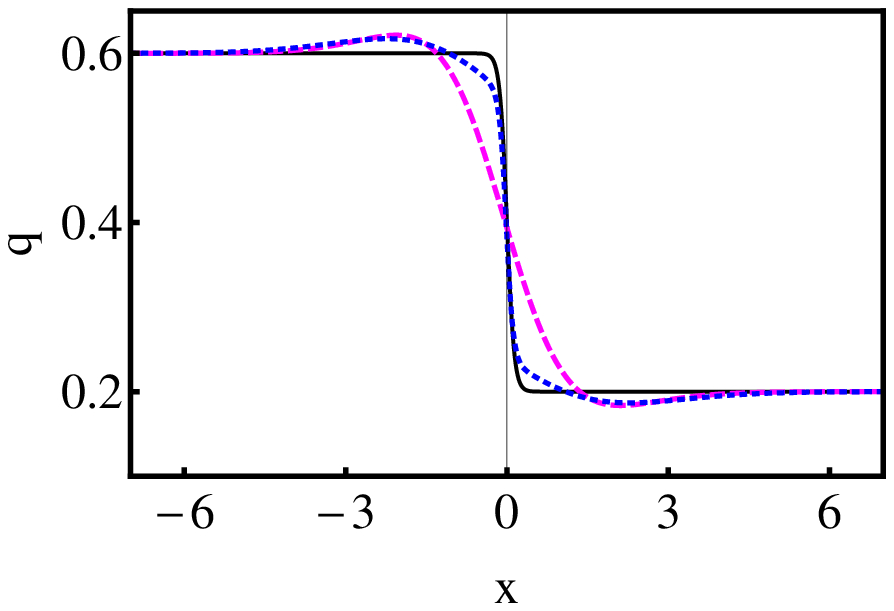}
\includegraphics[width=2.4 in,clip=]{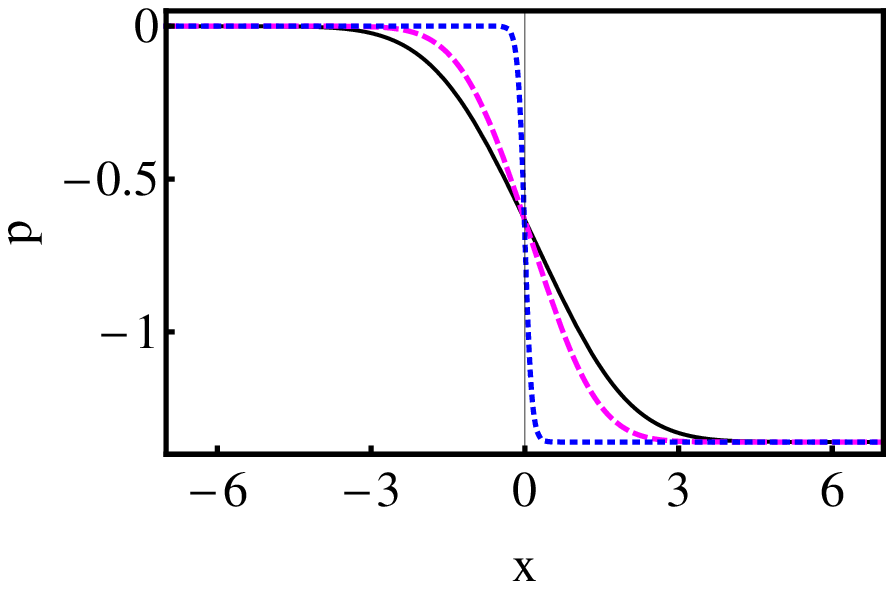}
\caption{(Color online) Same as in Figs. \ref{quenchedqp4} and \ref{qpminus4}, but with $\lambda=-1.36$ which corresponds
to $J\simeq 0$. The rescaled time moments are $t=0$ (solid line), $1/2$ (dashed line), and $1$ (dotted line).}
\label{quenchedqpzeroJ}
\end{figure}

Figures \ref{quenchedqp4} to \ref{quenchedqpzeroJ} show three examples  of numerically found optimal paths $q(x,t),\,p(x,t)$ for the SSEP in the quenched setting. Here
no analytic results are available beyond the small fluctuations, see Eq.~(\ref{variance_quenched}), except for the special cases of $\rho_-=\rho_+=1/2$ and $\rho_-=1,\,\rho_+=0$.  The case of $\lambda=4$ corresponds to a positive current a few times greater than the average current. Here too the optimal fluctuation grows
toward $t=1$ and facilitates transport of the material from left to right. Notice a striking similarity between the $p$-profiles for $\lambda=4$
in the annealed and quenched settings, for which we do not have a good explanation.

For $\lambda=-4$ the optimal fluctuation reverses the current compared with the hydrodynamic flow. Finally, for $\lambda \simeq -1.34$ the integrated current is equal to zero.
Here, after an initial release of material from left to right, the fluctuation pushes the material back. Still, as one can see from Fig.~\ref{quenchedqpzeroJ}, the final density profile $q(x,1)$ is different from the initial profile.

\begin{figure}[ht]
\includegraphics[width=2.6 in,clip=]{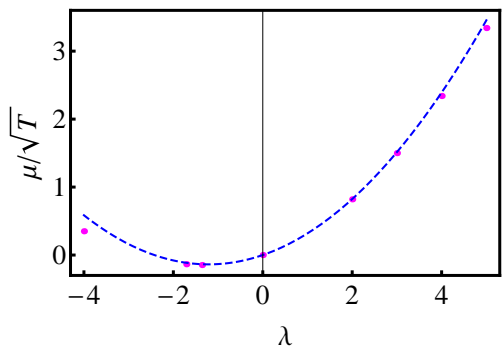}
\caption{(Color online). $\mu_{\text{quenched}}^{\text{SSEP}}/\sqrt{T}$ versus $\lambda$ for
$\rho_{-}=0.6$ and $\rho_{+}=0.2$. Circles: our numerical results. Dashed curve:
the two-cumulant approximation $C_1 \lambda + (1/2)\, V_{\text{quenched}} \lambda^2$  with $C_1$ from Eq.~(\ref{C1SSEP}) and $V_{\text{quenched}}$ from Eq.~(\ref{asymfullquen}).}
\label{muquen}
\end{figure}

Finally, Fig. \ref{muquen} shows the $\lambda$-dependence of $\mu_{\text{quenched}}^{\text{SSEP}}/\sqrt{T}$ that we found numerically
in a moderate range of $\lambda$ for $\rho_{-}=0.6$ and $\rho_{+}=0.2$. Also shown is the two-cumulant approximation
$\mu=C_1 \lambda+ (1/2)\, V_{\text{quenched}} \lambda^2$, with $C_1$ from Eq.~(\ref{C1SSEP}) and $V_{\text{quenched}}$ from Eq.~(\ref{asymfullquen}).
One can see that, as in the annealed setting, the two-cumulant approximation is quite accurate well beyond small $\lambda$. That is, both in the annealed and in the quenched settings, deviations of the probability distribution $P(J,T)$ from a Gaussian only occur in relatively far tails of the distribution.

\section{Concluding Remarks}
\label{concl}

We have investigated the long-time fluctuations of integrated current in diffusive lattice gases in one dimension, when the initial density is a step-like function. Our analysis relies on the macroscopic fluctuation theory (MFT) \cite{Bertini}, more precisely on its implementation \cite{DG2009b} which allows one to examine the fluctuations of the current in a non-stationary situation of a step-like initial density profile. For the quenched and annealed settings we have calculated the variance of the current fluctuations by developing a perturbation theory around the noiseless hydrodynamic solution. Our results for the variance hold for a whole family of lattice gas models which, at the coarse-grained level of the MFT, can be characterized by a constant diffusion coefficient and an arbitrary $\sigma(q)$.  Particular examples of our results include the variance for the non-interacting random walkers, for the symmetric exclusion process and for the  Kipnis-Marchioro-Presutti model. For the annealed setting these particular results agree with previous results. For the quenched setting these results are new.

We have also investigated numerically the regime of large deviations of the current.  Using the SSEP as a representative example, we have solved the MFT equations by using the Chernykh-Stepanov numerical iteration algorithm. We have found the optimal paths corresponding to an unusually large current, to a current that flows in the ``wrong'' direction, and to zero current. We have also computed numerically the large deviation function $\mu(\lambda)$ and observed that the two-cumulant approximation (and, correspondingly, the Gaussian asymptotic of the integrated current distribution) remain quite accurate well beyond the small-$\lambda$ regime. These numerical results are important because an analytical solution of the MFT equations, beyond small fluctuations, is presently unavailable except for non-interacting random walkers \cite{DG2009b}.

An important task for a future work is an asymptotic analysis of the MFT equations in the large current limits, $|J|\gg \langle J \rangle$.  Here the iteration algorithm may help in getting insight into the type of perturbation expansion needed for that.

\section*{Acknowledgments}

We are very grateful to Pavel Sasorov for discussions and collaboration at an earlier stage of this work.
B.M. was supported by the Israel Science Foundation (Grant No. 408/08), by the US-Israel Binational Science Foundation (Grant No. 2008075), and by the Condensed Matter Theory Visitors Program of Boston University's Physics Department.

\section*{Appendix A. Free energy from the MFT formalism}
\renewcommand{\theequation}{A\arabic{equation}}
\setcounter{equation}{0}
Here we show how Eq.~(\ref{FDT}) appears in the MFT formalism. (See also Ref. \cite{Tailleur} for a similar derivation in the particular case of the SSEP.) Consider a lattice gas in equilibrium at average density $\bar{\rho}=const$.
The probability of observing a given density profile $q(x)$ is described by the equilibrium Boltzmann-Gibbs distribution:
\begin{equation}\label{BG}
    -\ln {\cal P}[q(x)]
    \!\sim \!\!\int_{-\infty}^{\infty} dx \left[F(q(x))-F(\bar{\rho})-F^{\prime}(\bar{\rho})(q(x)-\bar{\rho})\right],
\end{equation}
see e.g. Ref. \cite{DG2009b}. In the MFT formalism, this expression should coincide with the mechanical action,
\begin{equation}
\label{action:app}
S_0 = \int_{-\infty}^0 dt \int_{-\infty}^\infty dx \left(p\,\partial_t q - \mathcal{H}\right),
\end{equation}
calculated along the activation trajectory of the MFT equations~(\ref{q:eq}) and (\ref{p:eq}) obeying the following boundary conditions in time:
$q(x,t=-\infty)=\bar{\rho}$ and $q(x, t=0)=q(x)$ \cite{Tailleur}.  We will now see how it happens, and how $F(q)$ emerges.

It is very simple to find the activation trajectory here because, in equilibrium, it coincides with a time-reversed relaxation trajectory. That is, the optimal fluctuation $q(x,t)$ must obey the time-reversed version of Eq.~(\ref{rho:eq}):
\begin{equation}
\label{reversed}
\partial_t q = -\partial_x \!\left[D(q) \,\partial_x q\right].
\end{equation}
Combining this equation with Eq.~(\ref{q:eq}), we obtain
\begin{equation}\label{der}
    \sigma(q) \partial_x p = 2 D(q) \partial_x q +f(t),
\end{equation}
where $f(t)=0$ because of the boundary conditions in $x$. Now we introduce function
$F(q)$ that satisfies Eq.~(\ref{FDT}). Integrating Eq.~(\ref{der}) over $x$ yields
\begin{equation}\label{p(q)}
    p=F^{\prime}(q) -F^{\prime}(\bar{\rho}),
\end{equation}
where we have demanded that $p=0$ at $q=\bar{\rho}$. A straightforward algebra shows that $p=F^{\prime}(q)-F^{\prime}(\bar{\rho})$ also solves Eq.~(\ref{p:eq}).
This local relation between $p$ and $q$ implies complete integrability of the MFT problem for equilibrium. Furthermore,
here $\mathcal{H}=0$, and the action $S_0$ becomes
\begin{eqnarray}
  S_0 &=& \int_{-\infty}^0 dt \int_{-\infty}^\infty dx\, p\,\partial_t q \nonumber \\
  &=&\int_{-\infty}^\infty dx \int_{-\infty}^0 dt \left[F^{\prime}(q) -F^{\prime}(\bar{\rho})\right]\partial_t q
\end{eqnarray}
which, upon integration over time, yields Eq.~(\ref{BG}) as expected.

In non-equilibrium situations the local relation $p=F^{\prime}(q)-F^{\prime}(\bar{\rho})$ breaks down, as it does not satisfy some or all of the boundary conditions. As a result, the MFT equations become, in general, non-integrable, whereas the mechanical action
explicitly depends on the system dynamics at intermediate times.

\section*{Appendix B. Quenched variance from fluctuating hydrodynamics}
\renewcommand{\theequation}{B\arabic{equation}}
\setcounter{equation}{0}

Equation~(\ref{variance_quenched}) for the variance in the quenched setting can be also obtained in the framework of fluctuating hydrodynamics, by solving a linearized Langevin equation. Once $D=1$ and $\sigma(\rho)$ are known, the Langevin equation for the fluctuating particle density field $q(x,t)$ can be written as \cite{Spohn,KL99}
\begin{equation}
\label{Langevin}
     \partial_t q = \partial_{xx} q+\partial_x \left[\sqrt{\sigma(q)} \,\xi(x,t)\right].
\end{equation}
Here $\xi(x,t)$ is a zero-average Gaussian noise, delta-correlated in space and in time:
\begin{equation}
\left\langle \xi(x,t)\xi(x_1,t_1)\right\rangle=\delta(x-x_1)\, \delta(t-t_1),
\label{deltacorr}
\end{equation}
and the brackets denote ensemble averaging. Linearizing the Langevin equation \eqref{Langevin} around the hydrodynamic solution \eqref{q_0}, we write $q=\rho+q_1$ where $q_1\ll \rho$. This yields
\begin{equation}
\label{Langlin}
     \partial_t q_1 - \partial_{xx} q_1=\partial_x \left[\sqrt{\sigma(\rho)} \,\xi(x,t)\right],
\end{equation}
Plugging $q_1=\partial_x \phi$, we can rewrite this equation as
\begin{equation}\label{Langlin1}
     \partial_t \phi - \partial_{xx} \phi=\sqrt{\sigma(\rho)} \,\xi(x,t).
\end{equation}
The solution is
\begin{eqnarray}
  \!\!\!\phi(x,t) &=& \int_0^t dt_1\int_{-\infty}^{\infty}dy\,\frac{\sqrt{\sigma(\rho)} \,\xi(y,t_1)}
  {\sqrt{4 \pi (t-t_1)}}\,e^{-\frac{(x-y)^2}{4 (t-t_1)}}. \label{psi}
\end{eqnarray}
The current $J(T)$ from Eq.~(\ref{current0}) can be represented as
$$
J(T)=\langle J \rangle +\int_0^{\infty} q_1(x,T)\,dx.
$$
The variance of the current is, therefore,
\begin{eqnarray}
\langle J^2\rangle_c &=& \left\langle \int_0^{\infty} dx \int_0^{\infty} dy \,q_1(x,T) q_1(y,T) \right\rangle \nonumber \\
   &=& \left\langle  \int_0^{\infty} dx \int_0^{\infty} dy \, \partial_x \phi(x,T) \,\partial_{y} \phi(y,T) \right\rangle \nonumber \\
   &=& \langle \phi^2 (0,T)\rangle.
   \label{var1}
\end{eqnarray}
Combining Eqs.~(\ref{psi}) and (\ref{var1}), we obtain
\begin{widetext}
\begin{eqnarray}
  \langle J^2\rangle_c  &=& \frac{1}{4\pi}\int_0^T dt_1\int_0^T dt_2 \int_{-\infty}^{\infty} dx
  \int_{-\infty}^{\infty} dy \,e^{-\frac{x^2}{4 (T-t_1)}-\frac{y^2}{4 (T-t_2)}}
  \sqrt{\sigma[\rho(x,t_1)]\,\sigma[\rho(y,t_2)]}\,\langle \xi(x,t_1) \xi(y,t_2) \rangle \nonumber \\
  &=&\frac{1}{4\pi}\int_0^T dt_1\int_0^T dt_2 \int_{-\infty}^{\infty} dx
  \int_{-\infty}^{\infty} dy \,e^{-\frac{x^2}{4 (T-t_1)}-\frac{y^2}{4 (T-t_2)}}
  \sqrt{\sigma[\rho(x,t_1)]\,\sigma[\rho(y,t_2)]}\,\delta(x-y) \delta(t_1-t_2) \nonumber \\
  &=&\frac{1}{4\pi} \int_0^T dt \int_{-\infty}^{\infty} dx \,\frac{e^{-\frac{x^2}{2(T-t)}}}{T-t}\,\sigma(\rho)\label{var2}
\end{eqnarray}
\end{widetext}
which coincides with Eq.~(\ref{variance}) obtained, in the quenched setting, from the MFT.


\begin{thebibliography}{99}

\bibitem{LLStat}
    L. D. Landau and E. M. Lifshitz, {\it Statistical Physics} (New York: Pergamon Press, 1980).

\bibitem{Spohn}
     H. Spohn, {\it Large Scale Dynamics of Interacting Particles}
    (New York: Springer-Verlag, 1991).

\bibitem{L99}
    T. M. Liggett, {\it Stochastic Interacting Systems: Contact, Voter, and Exclusion Processes}
    (Springer, New York, 1999).

\bibitem{KL99}
     C. Kipnis and C. Landim, {\it Scaling Limits of Interacting Particle Systems}
     (Springer, New York,  1999).

\bibitem{SZ95}
     B. Schmittmann and R. K. P. Zia, \textit{Statistical Mechanics of Driven Diffusive Systems},
     in: {\it Phase Transitions and Critical Phenomena}, Vol.\ 17, eds.\ C. Domb and J. L. Lebowitz
     (Academic Press, London, 1995).

\bibitem{D98} B. Derrida, Phys.\ Rep.\ {\bf 301}, 65 (1998).

\bibitem{S00} G. Sch\"utz, \textit{Exactly Solvable Models for Many-Body Systems
  Far From Equilibrium}, in {\it Phase Transitions and Critical Phenomena},
  Vol.\ 19, eds.\ C. Domb and J. L. Lebowitz (Academic Press, London, 2000).

\bibitem{BE07} R. A. Blythe and M. R. Evans, J. Phys.\ A {\bf 40}, R333
  (2007).

\bibitem{KRB10}
     P. L. Krapivsky, S. Redner, and E. Ben-Naim,
     {\it A Kinetic View of Statistical Physics} (Cambridge University Press, Cambridge,  2010).

\bibitem{NESS}
     H. Touchette, Phys. Rep. {\bf 478}, 1 (2009).

\bibitem{D07}
     B. Derrida, J. Stat. Mech. P07023 (2007).

\bibitem{Jona}
    G. Jona-Lasinio, Prog. Theor. Phys. Suppl. {\textbf 184}, 262 (2010).

\bibitem{Bertini}
    L. Bertini, A. De Sole, D. Gabrielli, G. Jona-Lasinio, and C. Landim,
    Phys. Rev. Lett. {\textbf 87}, 040601 (2001);  {\it ibid} {\textbf 94}, 030601 (2005);
    J. Stat. Phys. {\textbf 123}, 237 (2006); {\it ibid} {\textbf 135}, 857 (2009);
    J. Stat. Mech. (2007) P07014.

\bibitem{Tailleur}
    J. Tailleur, J. Kurchan, and V. Lecomte, Phys. Rev. Lett. \textbf{99}, 150602 (2007);
    J. Phys. A {\bf 41}, 505001 (2008).

\bibitem{FW84}
    M. I. Freidlin and A. D. Wentzell, {\it Random Perturbations of Dynamical Systems}
    (New York: Springer-Verlag, 1984).

\bibitem{MSR}
     P. C. Martin, E. D. Siggia, and H. A. Rose, Phys. Rev. A \textbf{8}, 423 (1973).

\bibitem{EK}
    V. Elgart and A. Kamenev, Phys. Rev. E \textbf{70}, 041106 (2004).

\bibitem{MS2011}
    B. Meerson and P. V. Sasorov, Phys Rev. E \textbf{83}, 011129 (2011).

\bibitem{MSfronts}
    B. Meerson, P. V. Sasorov, and Y. Kaplan, Phys Rev. E \textbf{84}, 011147 (2011);
    B. Meerson and P. V. Sasorov, Phys Rev. E \textbf{84}, 030101(R) (2011).

\bibitem{Appert-Rolland}
    C. Appert-Rolland, B. Derrida, V. Lecomte, and F. van Wijland, Phys. Rev. E \textbf{78}, 021122 (2008).

\bibitem{battery}
    T. Bodineau,  B. Derrida, and J.L. Lebowitz, J. Stat. Phys. \textbf{140}, 648 (2010).

\bibitem{van}
    V. Lecomte, J. P. Garrahan, and F. van Wijland,  J. Phys. A: Math. Theor. \textbf{45}, 175001  (2012).

\bibitem{DL}
    B. Derrida, J. L. Lebowitz, and E. R. Speer,  Phys. Rev. Lett. {\bf 87}, 150601 (2001);
    J. Stat. Phys. {\bf 107}, 599 (2002).

\bibitem{van_1}
    F. van Wijland and Z. Racz,  J. Stat. Phys. {\bf 118}, 27 (2005).

\bibitem{Rakos}
    R. J. Harris, A. R\'akos, and G. Sch\"utz, J. Stat. Mech. P08003 (2005).

\bibitem{KM}
    S. Prolhac and K. Mallick, J. Phys. A {\bf 41}, 175002 (2008); {\em ibid} {\bf 42}, 175001 (2009).


\bibitem{kurchan}
    C. Giardina, J. Kurchan and L. Peliti, Phys. Rev. Lett. {\bf 96}, 120603 (2006);  C. Giardina, J. Kurchan, V. Lecomte, and
    J. Tailleur,  J. Stat. Phys. {\bf 145}, 787 (2011).

\bibitem{pablo}
    P. I. Hurtado, C. P\'erez-Espigares, J. J. del Pozo and P. L. Garrido,
    Proc. Natl. Acad. Sci. USA {\bf 108}, 7704 (2011);
    P. I. Hurtado and P. L. Garrido, Phys. Rev. Lett. {\bf 107}, 180601 (2011);
    A. Prados, A. Lasanta and P. I. Hurtado, Phys. Rev. Lett. {\bf 107}, 140601 (2011).



\bibitem{vander}
    M. Gorissen, J. Hooyberghs, and C. Vanderzande, Phys. Rev. E {\bf 79}, 020101(R) (2009).

\bibitem{DG2009b}
    B. Derrida and A. Gerschenfeld, J. Stat. Phys. \textbf{137}, 978 (2009).

\bibitem{Stepanov}
    A. I. Chernykh and M. G. Stepanov, Phys. Rev. E \textbf{64}, 026306 (2001).

\bibitem{Bunin}
     G. Bunin, Y. Kafri, and D. Podolsky, EPL {\bf 99}, 20002 (2012).

\bibitem{KMP}
     C. Kipnis, C. Marchioro, and E. Presutti, J. Stat. Phys. {\textbf 27}, 65 (1982).

\bibitem{BGL}
    L. Bertini, D. Gabrielli,  and J. L. Lebowitz,  J. Stat. Phys. {\bf 121}, 843 (2005).

\bibitem{van_2}
    A. Imparato, V. Lecomte, and F. van Wijland,  Phys. Rev. E {\bf 80}, 011131 (2009).

\bibitem{DG2009a}
    B. Derrida and A. Gerschenfeld, J. Stat. Phys. \textbf{136}, 1 (2009).

\bibitem{PS02}
    M. Pr\"ahofer and H. Spohn, in: {\it In and Out of Equilibrium},
    ed.\ V. Sidoravicious (Birkh\"auser, Basel, 2002).

\bibitem{varadhan} S. Sethuraman and S.R.S. Varadhan, arXiv:1101.1479.

\bibitem{previous} This numerical instability was not observed when the same iteration algorithm was used in situations when, in addition
to diffusive transport, there are on-site reactions among particles \cite{EK,MS2011}.

\end{thebibliography}
\end{document}